# Title: Engineering Phonon Polaritons in van der Waals Heterostructures to Enhance In-Plane Optical Anisotropy

**Authors:** Kundan Chaudhary[1‡], Michele Tamagnone[1‡], Mehdi Rezaee[1,2‡], D. Kwabena Bediako[3], Antonio Ambrosio[4], Philip Kim[3], Federico Capasso[1*]

**Affiliations:**

[1]Harvard John A. Paulson School of Engineering and Applied Sciences, Harvard University, Cambridge, MA 02138, USA.
[2]Department of Electrical Engineering, Howard University, Washington, DC 20059, USA.
[3]Department of Physics, Harvard University, Cambridge, MA 02138, USA.
[4]Center for Nanoscale Systems, Harvard University, Cambridge, MA 02138, USA.

*Correspondence to: capasso@seas.harvard.edu

‡These authors contributed equally to this work.

**Abstract**: Van der Waals heterostructures assembled from layers of 2D materials have attracted considerable interest due to their novel optical and electrical properties. Here we report a scattering-type scanning near field optical microscopy study of hexagonal boron nitride on black phosphorous (h-BN/BP) heterostructures, demonstrating the first direct observation of in-plane anisotropic phonon polariton modes in vdW heterostructures. Strikingly, the measured in-plane optical anisotropy along armchair and zigzag crystal axes exceeds the ratio of refractive indices of BP in the x-y plane. We explain that this enhancement is due to the high confinement of the phonon polaritons in h-BN. We observe a maximum in-plane optical anisotropy of $\alpha_{max} = 1.25$ in the 1405-1440 cm$^{-1}$ frequency spectrum. These results provide new insights on the behavior of polaritons in vdW heterostructures, and the observed anisotropy enhancement paves the way to novel nanophotonic devices and to a new way to characterize optical anisotropy in thin films.

**Introduction**

Phonon polaritons due to the coupling of photons and optical phonons in polar dielectrics such as hexagonal boron nitride (h-BN) have attracted much interest in the two-dimensional (2D) materials research-field (*1-10*). Specifically, h-BN has light constituent elements which lead to strong phonon resonances at mid-infrared (mid-IR) frequencies (*10*). Such h-BN based polaritonic systems exhibit confinement of light below the diffraction limit (*11*), propagation with ultra-low losses (*3*), hyperbolic properties (*5, 10-15*), and tunable response by changing the thickness of the samples (*10*) at mid-IR wavelengths. However, it is noteworthy that in the experiments performed so far, the substrates were optically isotropic. The effect of the substrate in-plane optical anisotropy ($\varepsilon_x \neq \varepsilon_y$, where $\varepsilon_x$ and $\varepsilon_y$ are the relative permittivities along *x* and *y* axes) has not yet been investigated. It is therefore important to gain new insights on the interaction of polaritons with in-plane optically anisotropic substrates. In this work, we report on a scattering-type scanning near field optical microscopy (s-SNOM) (*16, 17*) study of h-BN on black phosphorus (h-BN/BP) heterostructures, demonstrating the first direct observation of in-plane anisotropic phonon polariton waveguide modes supported by h-BN placed on BP, an in-plane anisotropic van der Waals (vdW) material (*18-20*). Remarkably, as explained later, the maximum anisotropy of the phonon polariton propagation along *x* and *y* axes of our heterostructure exceeds the anisotropy of the vdW layers used to create it.

**Results**

Our samples consist of a thin (40 nm) h-BN flake placed on BP on a SiO$_2$/Si substrate and then patterned into disc shapes (see Supplementary Figures 1-3). Due to their rotational symmetry any anisotropy in the polariton propagation would be clearly visible. In addition, and for the same reason, no careful alignment is needed between the patterns and the BP optical axes at the stage of device fabrication (see Supplementary Method 1). The phonon polaritons in h-BN and optical properties of BP have been explored in several works (*10, 21*) and are briefly recalled here. h-BN (Figure 1a) is a uniaxial anisotropic (or birefringent) material characterized by an in-plane relative dielectric permittivity $\varepsilon_\perp$ (in the *x* and *y* directions of Figure 1a) different from the out-of-plane one $\varepsilon_\parallel$, (in the *z* direction). The polarization of the boron and nitrogen bond allows for the coupling of optical phonons with incident infrared light, resulting in phonon polaritons. Because the unit cell contains two atoms, there are three optical phonon modes, two in-plane (degenerate due to the in-plane isotropy of h-BN) with resonance at 1370 cm$^{-1}$ and one out-of-plane, with lower resonance frequency 780 cm$^{-1}$. The presence of the polaritons causes a negative permittivity in two different frequency bands, called reststrahlen bands, as shown in Figure 1b (first reststrahlen band, RS1: 780-830 cm$^{-1}$; second reststrahlen band, RS2: 1370-1610 cm$^{-1}$) (*7, 10*). In the RS1 band, the relative out-of-plane permittivity, $\varepsilon_\parallel$, is negative, whereas in the RS2 band, considered in this work, the relative in-plane permittivity, $\varepsilon_\perp$, is negative. As a result of these optical properties, h-BN supports highly confined guided transverse magnetic (TM) modes in these two bands (*7, 10*). It is instructive to recall the propagation of TM plane waves (i.e., p-polarized) in h-BN. In this uniaxial material, the dispersion reads:

$$\frac{k_\perp^2}{\varepsilon_\parallel} + \frac{k_\parallel^2}{\varepsilon_\perp} = k_0 = \frac{\omega^2}{c^2} \qquad (1)$$

where $k_\perp$ and $k_\parallel$ are the in-plane and out-of-plane component of the wavevector, $k_0$ is the wavevector in vacuum at the frequency $\omega$ and $c$ is the speed of light. Equation (1) is represented graphically in Figure 1c. The negative out-of-plane or in-plane permittivity in the reststrahlen bands enables the propagation of hyperbolic polaritons, which can have arbitrarily high wavevectors. This allows even thin flakes of h-BN to support an infinite number of guided modes theoretically (*10*). Unlike h-BN, BP is a non-polar dielectric and hence does not exhibit phonon polaritons. Whereas highly doped (or gated) BP supports plasmon polaritons (*22*), the undoped material used in this work does not (Table 1) and behaves as an anisotropic biaxial dielectric (with relative permittivities, $\varepsilon_x \neq \varepsilon_y \neq \varepsilon_z$, and $n_{x,y,z} = \sqrt{\varepsilon_{x,y,z}}$). The puckered orthorhombic geometry of BP (*18-20*) results in in-plane optical anisotropy with crystal axes (armchair and zigzag) exhibiting different refractive indices (Figure 1d and 1e). The in-plane optical anisotropy is prominent in the entire optical spectrum of interest from ultraviolet to infrared (*21-23*), which allows the determination of the crystal axes using angle-resolved polarized Raman (ARPR) spectroscopy (*24*) (Supplementary Figure 1a,b). In addition, polarized light microscopy (Supplementary Figure 2) is used to identify the crystal axes of BP. Figure 1f illustrates the electric field lines of the excited fundamental guided phonon polariton mode supported by h-BN. Here, the field lines are influenced by the presence of BP and its in-plane anisotropy, which causes the in-plane guided polariton modes to be in-plane anisotropic as well. More precisely, for a constant illumination wavelength, the polaritons will propagate with different effective indices along different in-plane crystal axes of BP. Figure 1g qualitatively compares contour plot of dispersion relation at a given frequency of isotropic phonon polariton modes in h-BN to that of anisotropic phonon polariton modes in h-BN/BP heterostructure. The dispersion relation of anisotropic phonon polaritons in h-BN/BP heterostructure is approximated by an ellipse in the $k_x$-$k_y$ plane as opposed to a circle for the case of isotropic phonon polaritons in h-BN (*14*). We then define the polariton anisotropy $\alpha$ as:

$$\alpha = \frac{n_{eff,x}}{n_{eff,y}} \tag{2}$$

where $n_{eff,x}$ and $n_{eff,y}$ are the effective indices of the guided polariton modes in the *x* (armchair) and *y* (zigzag) directions, respectively. Notably, the polariton anisotropy, $\alpha$, defined with the effective indices of the modes in the h-BN/BP heterostructure, can be larger than the ratio ($\alpha'$) of the refractive indices of BP in the x-y plane:

$$\alpha' = \frac{n_x}{n_y} \qquad (3)$$

The guided modes in this heterostructure therefore exhibit an enhanced anisotropy with respect to BP alone. Figure 2a shows the effect of thickness of BP on the calculated (using Lumerical Mode Solutions) effective indices of the phonon polariton modes along the armchair and zigzag axes of BP. In the frequency spectrum 1405-1440 cm$^{-1}$ there is a clear evidence of anisotropy in the phonon polariton propagation along the armchair and the zigzag axes given by the difference in their effective indices. The effective indices along both the armchair and the zigzag direction increase with frequency.

Figure 2b demonstrates the in-plane anisotropy, $\alpha$, of phonon polariton propagation in h-BN/BP heterostructures (Figure 2c). The dashed-purple line represents the ratio of refractive indices of BP in the x-y plane, $\alpha'$. As mentioned earlier, there are wide spectral ranges where $\alpha$ exceeds $\alpha'$. This effect can be understood as follows: let us consider a polariton propagating along the zigzag direction, for a BP thickness such that polariton fields are partially inside BP. When propagation along the armchair axis is considered, the electric field (which is partially polarized along the wavevector) interacts with a higher permittivity which slows down the wave propagation. However, this also implies that the field is now more confined, i.e. a larger fraction of it is inside h-BN. This additional effect further reduces the phase velocity, enhancing the anisotropy. As shown in Figure 2b, for each frequency there is an optimum thickness of BP which maximizes the in-plane anisotropy. Similarly, for each thickness of BP, there is an optimum frequency which maximizes the in-plane optical anisotropy. In the large frequency limit (i.e., 1440 cm$^{-1}$), the in-plane optical anisotropy decreases with increasing thickness of BP.

Figure 2d shows the electric field profile of h-BN on a SiO$_2$/Si substrate where the electric field is mostly confined in h-BN and exponentially decays away from the h-BN surfaces in air and SiO$_2$. The confinement of polariton mode in h-BN along x-y plane increases with increasing BP thickness (Figure 2e,g,i), which allows for weaker interaction with the in-plane anisotropic refractive indices of BP. For 40 nm BP, the polariton mode is less confined in h-BN which allows for a stronger interaction with BP and leads to a higher difference in phonon polariton propagation along armchair and zigzag crystal axes. Similarly, the dispersion behavior of anisotropy can be explained by the fact that mode confinement in h-BN, for a given thickness of BP, is a function of frequency where the modes are highly confined for larger frequencies (Figure 2a). Similarly, Figure 2f,h,j demonstrates the electric field profile along out-of-plane axis of the h-BN/BP heterostructure.

In order to probe the fundamental guided mode of the anisotropic phonon polaritons in h-BN/BP heterostructure, we used s-SNOM which allows imaging of guided modes in the mid-IR spectrum. We used a quantum cascade laser (QCL) as a source focused by a parabolic mirror to a region of sample and to the probe (a PtIr-coated atomic force microscopy (AFM) tip). Scattering of the laser beam at the tip (diameter: ~20 nm) provides wavevector matching and excites the phonon polariton modes in h-BN (Figure 3a-c) in the RS2 band of h-BN. An interferometer and pseudo-heterodyne detection scheme are used to extract the amplitude and phase of the phonon polaritons in the h-

BN/BP heterostructure (*16*). The s-SNOM produces images which include both the amplitude and the phase of the scattered field at each pixel of the scan (raw scans are reported in Supplementary Figure 4), which can be obtained by using several of the harmonics of the pseudo-heterodyne detection. As explained in our previous works (*6, 25*), the complex-valued images are not a simple representation of the near fields of the structure, but rather they are a superposition of several contributions. Each contribution is associated to a particular path that the light can follow when interacting with the tip-sample system, and all the contributions are added together in the final detected image. The contributions as the final image, are represented by complex-valued images. For samples supporting no guided waves or standing resonances, only one path is relevant, which we call *material* contribution, and it is associated with the light path from the interferometer to the tip, which enhances light intensity on the material directly underneath it and then scatters it back to the interferometer (Figure 3a). This contribution is affected by the local material polarizability directly below the AFM tip.

When guided waves are supported, other contributions are possible. However, no guided wave can be detected unless there is an edge or any other scatterer on the sample. In our case, the samples are discs, the edge of which allows the detection of guided waves. Two possible paths contribute to the detection of the guided modes in the sample, and they are associated to the *roundtrip* and *direct* contributions. In the *roundtrip* contribution, the tip couples the QCL light into the guided mode, which is reflected by the edge of the sample and then is scattered back to the detector by the tip (Figure 3b). In the *direct* contribution, the light from the QCL is scattered by the edge into the guided mode, and then the tip scatters it back to the detector (Figure 3c). Hence, from the raw images (Supplementary Figure 4), we first remove the *material* contribution and then we separate the *direct* and *roundtrip* components exploiting the rotational symmetry of the system, as explained in Supplementary Method 1 and Supplementary Figure 5.

The roundtrip contribution is unaffected by possible misalignments (see Supplementary Method 1) of the laser beam with respect to the crystal axes of BP, and hence it is used for further analysis of the in-plane anisotropy of the polaritons in the h-BN/BP heterostructures (Figures 3 e-l). For a reference sample without BP (Supplementary Figure 5a), circular fringes are observed as expected from previous experiments (*14*). The presence of BP affects the phonon polaritons in h-BN and results in an anisotropic propagation given by the elliptical fringes with increased ellipticity towards the center of the disc. The orientation of the major and minor axes of ellipses represent the crystal axes of BP and matches that of the crystal axes measured using ARPR spectroscopy (Supplementary Figure 1a,b). Figure 3e-h shows the frequency dependence of the roundtrip contribution of the phonon polaritons in 40 nm h-BN/40 nm BP heterostructure, whereas Figure 3i-l shows the frequency dependence of phonon polaritons in 40 nm h-BN/250 nm BP. Elliptical fringes are observed for both 40 nm and 250 nm BP heterostructures, and the effective indices can be extracted from the fringe spacing measured as illustrated in Figure 3d. More precisely, the spacing between fringes is half of the guided wavelength of the mode since the waves propagate back and forth along the distance from the tip to the edge. The increased ellipticity of the fringes towards the center can be represented by the expressions: $a = \frac{m\lambda_{g,x}}{2}$ and $b = \frac{m\lambda_{g,y}}{2}$, where *a* and *b* are the distances from the first fringe (*m = 0*) to the $m^{th}$ fringe along *x* and *y* crystal axes, respectively, *m* is a positive integer, $\lambda_{g,x}$ and $\lambda_{g,y}$ are the guided polariton wavelengths along *x* and *y* crystal axes, respectively. The effective index increases with increasing frequency (Figure 3m) as expected for phonon polaritons in h-BN. There is a good agreement between the calculated effective indices and the measured effective indices using s-SNOM near-field images despite no fit was used. Instead, we use a theoretical prediction, starting from the measured thickness of the sample and from calculated values of the refractive indices of BP in ref. 21.

Thicker BP (in the semi-infinite limit) allows for a larger effective index, but the in-plane anisotropy is weaker. Hence, the heterostructure results in fringes with lower ellipticity away from the sample center. For 40 nm BP, the in-plane anisotropy peaks at 1420 cm$^{-1}$ (Figure 3n). On the other hand, the anisotropy decreases monotonically with frequency for 250 nm BP. There is an excellent agreement between the calculated and the experimental in-plane anisotropy. The maximum possible anisotropy given by the ratio of refractive indices of BP in the x-y plane (*21*) is $\alpha' = 1.13$, but the calculated and experimental values show a much higher maximum anisotropy of $\alpha_{max} = 1.25$ for 40 nm BP at 1420 cm$^{-1}$. This confirms the in-plane anisotropy enhancement (and for the first time) in h-BN/BP heterostructure described above. Similarly, in-plane anisotropy can be probed in the first reststrahlen band (RS1) of h-BN, but QCLs for this range is not available (*6*). We verified that rotating by 90 degrees the sample causes a rotation of the fringes orientation, as expected for an anisotropic sample (Supplementary Figure 5). Similar results are observed for all the harmonics of the pseudo-heterodyne detection used in our s-SNOM system (Supplementary Figure 6).

**Discussion**

We expect that the ability of engineering these deeply subwavelength modes will have important applications in mid-IR nanophotonics from probing the in-plane optical anisotropy of other vdW materials and heterostructures in mid-IR, which is clearly not available,to new approaches for designing vertically stacked heterostructures with extreme light confinement and tailored optical properties. For example, the heterostructure comprising of BP/h-BN/BP can lead to even higher in-plane optical anisotropy, up to $\alpha = 1.5$ (Figure S7). Furthermore, our study can be extended to other heterostructures which are electrostatically gated to tune in-plane optically anisotropic substrates such as BP and ReS$_2$ (*26*) to result in tunable phonon polaritons supported by polar dielectric materials such as h-BN and SiC (*16, 17*), and tunable plasmon polaritons supported by graphene (*27, 28*).

**Materials and Methods**

**Sample Fabrication:** BP flakes are mechanically exfoliated on to a 300 nm SiO$_2$/Si substrate in the Ar glovebox which has less than 0.1 ppm of O$_2$ and H$_2$O to preserve the flakes from deterioration. h-BN flakes are similarly exfoliated on the substrate. The thickness of flakes was confirmed with AFM (Park AFM). We used dry transfer technique with polymer (poly carbonate) and polydimethylsiloxane (PDMS) stamp to fabricate h-BN/BP heterostructures in the Ar glovebox. Substrates are then coated with poly (methyl methacrylate) (PMMA) 950(A6) and exposed with an electron beam system with a dose of 450 μC/cm$^2$ using an accelerating voltage of 30 kV. After developing in methyl isobutyl ketone (MIBK) for 1 minute, reactive ion etching (RIE) using a mixture of CHF$_3$, Ar, and O$_2$ at flow 10/5/2 sccm, respectively, and a RF generator at 30 W for 3-6 minutes was subsequently used to shape the heterostructure into discs. After the etching process, PMMA was removed by acetone and rinsed with isopropyl alcohol (IPA) and dried with nitrogen.

**Angle-Resolved Polarized Raman Spectroscopy:** Angle-resolved polarized Raman spectroscopy was performed using a 532 nm laser in the Horiba system. The power incident on BP was kept below 2.5 mW to avoid sample damage. Parallel polarization was used to collect the Raman signals. BP samples were rotated 360° about the microscope optical axis in 36 steps (10°/step). The grating number of detector was set as 1800 and spectral range from 300 to 500 cm$^{-1}$. The acquisition time was set as 10 s for 3 times accumulation to minimize the laser damage.

**Numerical Simulations:** The 1-D solver (Lumerical Mode Solutions) with a mesh size of 1 nm was used to compute the fundamental mode profile and effective indices of the h-BN/BP vdW heterostructures for a range of frequencies in the RS2 band of h-BN. We calculated the effective indices for a range (30-250 nm) of thickness of BP with a given thickness of h-BN (40 nm). Here, h-BN was modeled as an anisotropic dielectric with its permittivity values obtained from the Lorentz model (*6, 10, 25*). Besides, BP was modeled as an anisotropic dielectric with its permittivity values obtained from Valagiannopoulos et al. (21).

**Scattering-type Scanning Near Field Optical Microscopy (s-SNOM):** The near field scans were obtained using a commercially available s-SNOM from NeaSpec (www.neaspec.com) which is based on a tapping-mode AFM. Quantum Cascade Laser (QCL) was used as a tunable mid-IR source from Daylight Solutions (www.daylightsolutions.com). An IR beam from a QCL was focused on to a PtIr coated Si tip (diameter: ~20 nm) to launch the phonon polaritons modes. The back-scattered signal was demodulated at the higher harmonics ($n \geq 2$), of the tapping frequency to reduce the background. Details of the separation of the *direct*, *roundtrip* and *material contrast* contributions are presented in the Supplementary Information.

**References**

1. Basov, D. N. *et al.* Polaritons in van der Waals materials. *Science* **354** (2016).

2. Low, T. *et al.* Polaritons in layered two-dimensional materials. *Nat. Mater.* **16,** 182–194 (2017).

3. Giles, A. J. *et al.* Ultralow-loss polaritons in isotopically pure boron nitride. *Nat. Mater.* **17,** 134–139 (2018).

4. Shi, Z. *et al.* Amplitude- and Phase-Resolved Nanospectral Imaging of Phonon Polaritons in Hexagonal Boron Nitride. *ACS Photonics* **2,** 790–796 (2015).

5. Li, P. *et al.* Infrared hyperbolic metasurface based on nanostructured van der Waals materials. *Science* **359,** 892–896 (2018).

6. Ambrosio, A. *et al.* Selective excitation and imaging of ultraslow phonon polaritons in thin hexagonal boron nitride crystals. *Light Sci. Appl.* **7**, 27 (2018).

7. Caldwell, J. D. *et al.* Low-loss, infrared and terahertz nanophotonics using surface phonon polaritons. *Nanophotonics* **4,** 44–68 (2015).

8. Yoxall, E. *et al.* Direct observation of ultraslow hyperbolic polariton propagation with negative phase velocity. *Nat. Photon.* **9,** 674–678 (2015).

9. Tsakmakidis, K. L. *et al.* Ultraslow waves on the nanoscale. *Science* **358** (2017).

10. Dai, S. *et al.* Tunable phonon polaritons in atomically thin van der Waals crystals of boron nitride. *Science* **343,** 1125–1130 (2014).

11. Caldwell, J. D. *et al.* Sub-diffractional volume-confined polaritons in the natural hyperbolic material hexagonal boron nitride. *Nat. Commun.* **5**, 221 (2014).

12. Li, P. *et al.* Hyperbolic phonon-polaritons in boron nitride for near-field optical imaging and focusing. *Nat. Commun.* **6,** 1–9 (2015).

13. Kumar, A. *et al.* Tunable light-matter interaction and the role of hyperbolicity in

graphene-hBN system. *Nano Lett.* **15,** 3172–3180 (2015).

14. Ambrosio, A. *et al.* Mechanical Detection and Imaging of Hyperbolic Phonon Polaritons in Hexagonal Boron Nitride. *ACS Nano* **11,** 8741–8746 (2017).

15. Duan, J. *et al.* Launching Phonon Polaritons by Natural Boron Nitride Wrinkles with Modifiable Dispersion by Dielectric Environments. *Adv. Mater.* **29,** 1–8 (2017).

16. Huber, A. *et al.* Near-field imaging of mid-infrared surface phonon polariton propagation. *Appl. Phys. Lett.* **87**, 081103 (2005).

17. Huber, A. J. *et al.* Local excitation and interference of surface phonon polaritons studied by near-field infrared microscopy. *J. Microsc.* **229,** 389–395 (2008).

18. Ling, X. *et al.* The renaissance of black phosphorus. *Proc. Natl. Acad. Sci.* **112,** 4523–4530 (2015).

19. Xia, F. *et al.* Rediscovering black phosphorus as an anisotropic layered material for optoelectronics and electronics. *Nat. Commun.* **5,** 1–6 (2014).

20. Xia, F. *et al.* A. Two-dimensional material nanophotonics. *Nat. Photon.* **8**, 899-907 (2014).

21. Valagiannopoulos, C. A. *et al.* Manipulating polarized light with a planar slab of Black Phosphorus. *J. Phys. Commun.* **1**, 045003 (2017).

22. Low, T. *et al.* Plasmons and screening in monolayer and multilayer black phosphorus. *Phys. Rev. Lett.* **113,** 3–7 (2014).

23. Asahina, H. *et al.* Band structure and optical properties of black phosphorus. *J. Phys. C: Solid State Phys.*, **17**, 1839-1852 (1984).

24. Wang, T. *et al.* Identifying the Crystalline Orientation of Black Phosphorus by Using Optothermal Raman Spectroscopy. *Angew. Chem. Int. Ed.* **54**, 2366-2369 (2015).

25. Tamagnone, M. e*t al.* Ultra-confined mid-infrared resonant phonon polaritons in van der Waals nanostructures. *Sci. Adv.* **4**, eaat7189 (2018).

26. Chenet, D. A. *et al.* In-Plane Anisotropy in Mono- and Few-Layer $ReS_2$ Probed by Raman Spectroscopy and Scanning Transmission Electron Microscopy. *Nano Lett.* **15,** 5667–5672 (2015).

27. Fei, Z. *et al.* Gate-tuning of graphene plasmons revealed by infrared nano-imaging. *Nature* **486,** 82–85 (2012).

28. Chen, J. *et al.* Optical nano-imaging of gate-tunable graphene plasmons. *Nature* **487,** 77–81 (2012).

**Acknowledgments:**

General: The authors warmly thank Dr. Marios Mattheakis for the fruitful discussions.

Funding: This work was supported by the NSF EFRI, award no. 1542807. This work was performed in part at the Center for Nanoscale Systems (CNS), a member of the National Nanotechnology Coordinated Infrastructure Network (NNCI), which is supported by the National Science Foundation under NSF award no. 1541959. M.T. acknowledges the support of the Swiss National Science Foundation (SNSF) grant no. 168545 and 177836. M.R. and D.K.B. acknowledge the support of the Science and Technology Center for Integrated Quantum Materials, NSF Grant No. DMR-1231319.

Author Contributions: K.C., M.T., and F.C. devised experiments. M.R. and D.K.B. fabricated the samples. K.C., M.T., and A.A. performed s-SNOM scans. K.C. and M.T. analyzed the experimental data. K.C., M.T., and F.C. prepared the manuscript with input from all authors. All authors contributed to discussions and manuscript revision.

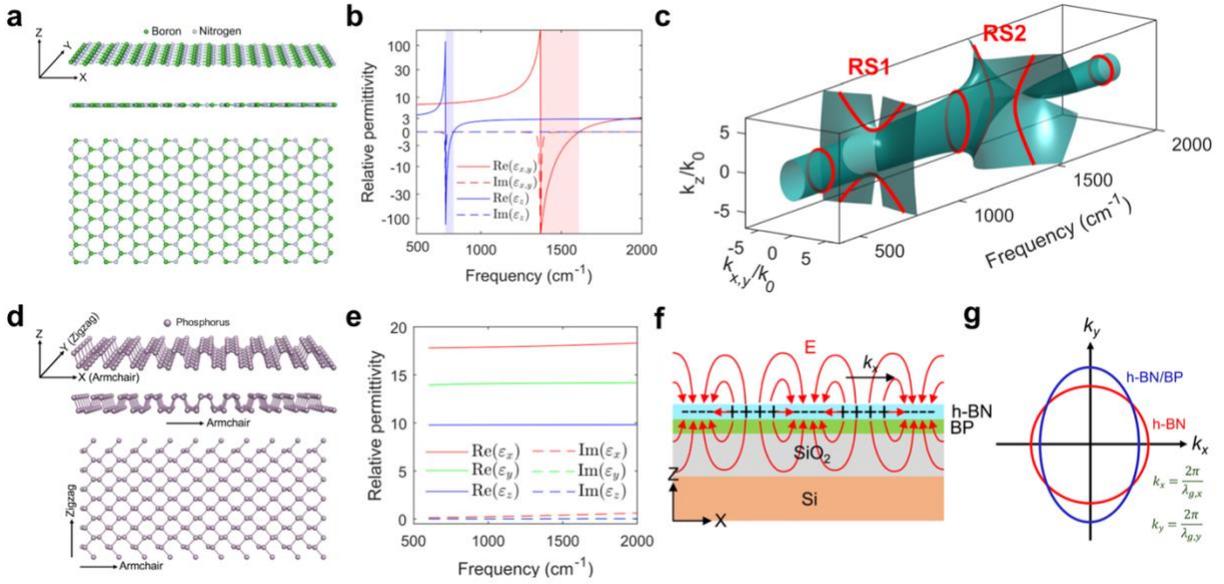

**Figure 1. Anisotropic phonon polariton propagation in h-BN/BP heterostructure. (a)** Schematic illustration of the crystal structure of hexagonal boron nitride (h-BN). A single layer is shown. **(b)** Relative permittivities of h-BN along different crystal axes showing hyperbolic behavior in two different frequency bands (highlighted in blue and red). **(c)** Schematic illustration of the hyperbolic dispersion relation of h-BN in the two different reststrahlen bands (RS1: 780-830 cm$^{-1}$, RS2: 1370-1610 cm$^{-1}$). **(d)** Schematic illustration of the puckered orthorhombic crystal structure of black phosphorus (BP). A single layer is shown. In the in-plane anisotropic geometry of BP, the slow axis is along the armchair direction whereas the zigzag direction represents the fast axis. **(e)** Relative permittivities of biaxial anisotropic BP along different crystal axes (*21*). **(f)** Electric field profile of polaritons in h-BN/BP heterostructure placed on SiO$_2$/Si substrate. The electric field lines here are due to coupling of mid-IR photons to in-plane optical phonons in the RS2 band of h-BN. **(g)** The presence of an in-plane anisotropic substrate such as BP affects the propagation of phonon polariton along armchair and zigzag crystal axes resulting in approximately elliptical dispersion relation in the $k_x$-$k_y$ plane in comparison to a circular dispersion relation for h-

BN on SiO$_2$/Si. $\lambda_{g,x}$ and $\lambda_{g,y}$ are the guided wavelengths of phonon polariton modes along $x$ and $y$ axes, respectively.

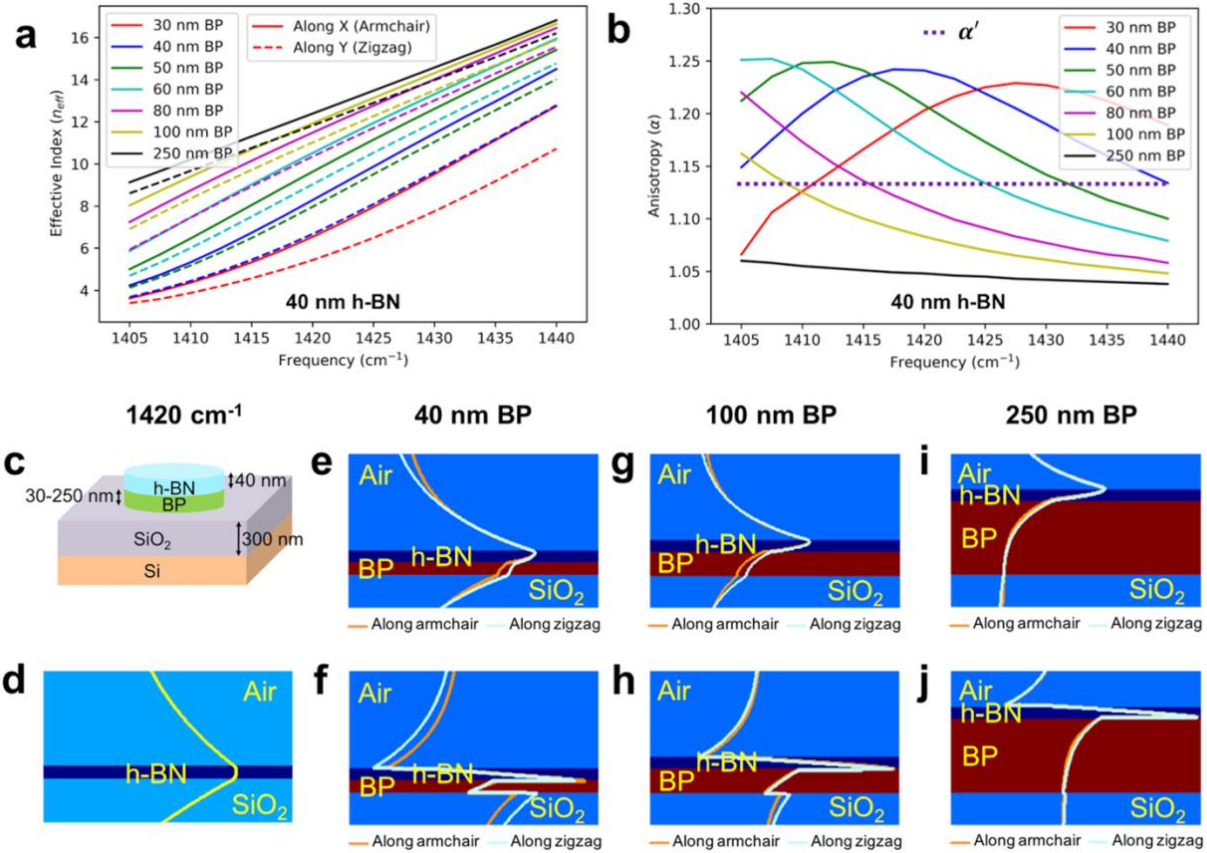

**Figure 2. Calculation of anisotropic dispersion relation of h-BN/BP heterostructure.** (**a**) For a given thickness (40 nm) of h-BN, the effective indices of phonon polariton modes along both armchair and zigzag crystal axes of BP increase monotonically with thickness of BP and frequencies in the range 1405-1440 cm$^{-1}$ due to increased mode confinement factor. (**b**) Calculated anisotropy values of the phonon polaritons for varying thickness of BP. For each frequency, there is an optimum thickness of BP which results in maximum anisotropy. Similarly, for each thickness of BP, there is an optimum frequency at which the anisotropy is maximum. The dashed-purple line represents the ratio of refractive indices of BP in the x-y plane. The calculated in-plane anisotropy of h-BN/BP heterostructure is larger ($\alpha_{max} = 1.25$) than the ratio of refractive indices of BP in the x-y plane ($\alpha' = 1.13$). (**c**) Schematic of h-BN/BP heterostructure used to calculate the effective indices in (**a**) and anisotropy in (**b**). For a fixed 40 nm thick h-BN, thickness of BP is varied

between 30 and 250 nm. **(d-j)** Electric field profiles at 1420 cm$^{-1}$. **(d)** Electric field profile with no BP. The mode is transverse magnetic (TM), and only the field component parallel to the propagation direction is plotted. Electric field is confined in h-BN demonstrating a guided polariton mode. The electric field intensity decays away from h-BN surfaces in air and SiO$_2$. **(e-f)** Electric field profile with 40 nm BP representing calculated in-plane and out-of-plane field components, respectively. The electric field confinement along *x* (armchair) is larger than *y* (zigzag) implying anisotropic effective indices along *x* and *y* crystal axes of BP. **(g-h)** Electric field profile with 100 nm BP representing in-plane and out-of-plane field components, respectively. The electric field confinement along both *x* and *y* are comparable implying lower anisotropy in comparison to the case with 40 nm BP. Higher confinement with 100 nm BP also demonstrates larger effective indices compared to the case with 40 nm BP. **(i-j)** Electric field profile with 250 nm BP representing in-plane and out-of-plane field components, respectively. Like the case with 100 nm BP, the electric field confinement along both *x* and *y* are comparable implying lower anisotropy in comparison to the case with 40 nm BP. Higher confinement in **(i-j)** also demonstrates larger effective indices compared to the case with 40 and 100 nm BP.

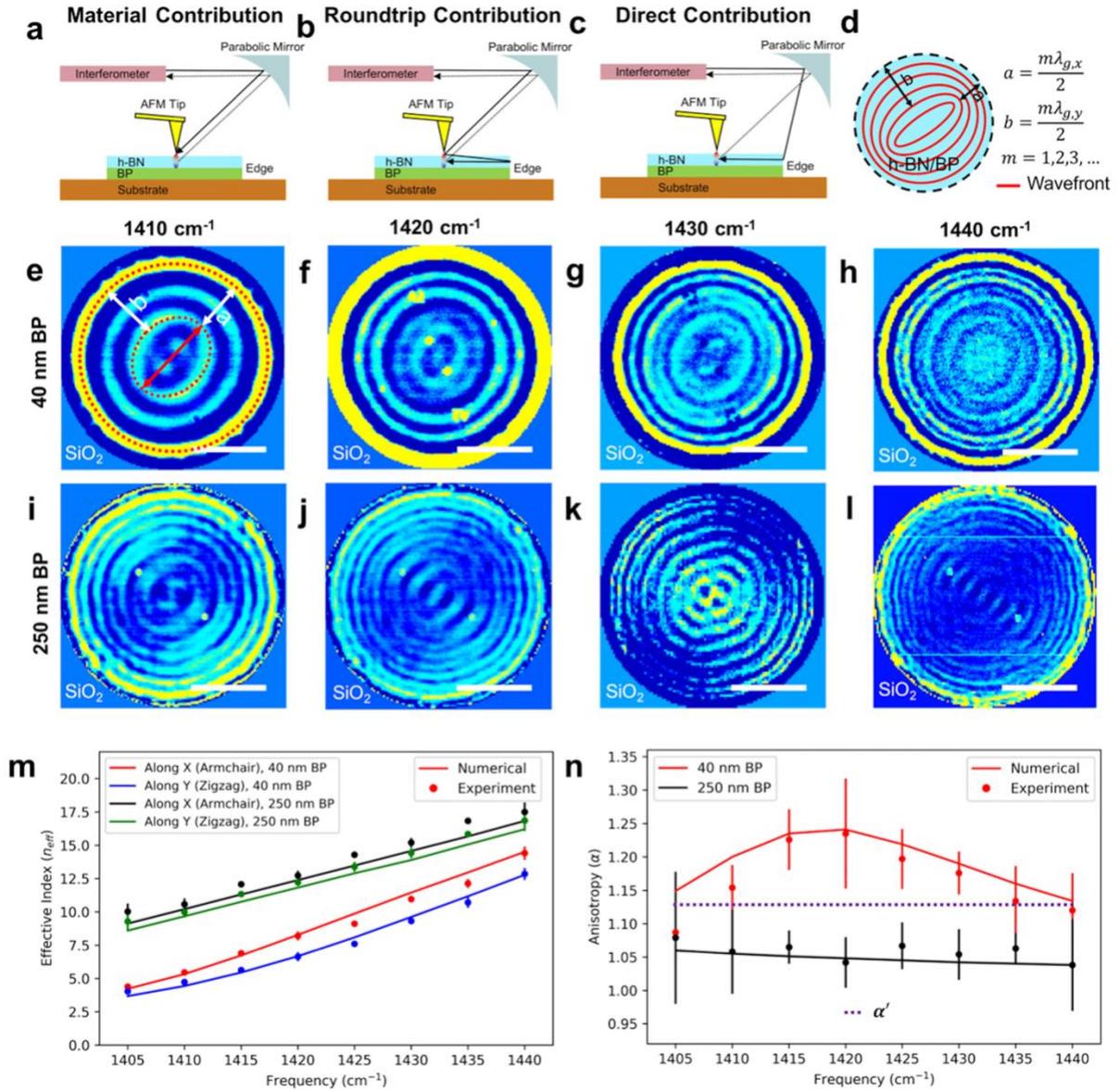

**Figure 3. Anisotropic dispersion relation of h-BN/BP heterostructure discs.** (**a**) Schematic illustration of *material* contribution. The incident IR beam goes to the tip, which focuses light on the material directly underneath and then scatters it back to the interferometer. (**b**) Schematic illustration of *roundtrip* contribution. The incident IR beam scatters off of the AFM tip, gets reflected from the edge of h-BN/BP heterostructure, and gets scattered again off of the AFM tip to the interferometer resulting in a round trip of the phonon polariton mode. (**c**) Schematic illustration of *direct* contribution. In this case the incident IR beam gets coupled to phonon polariton mode at the edge of the h-BN/BP heterostructure and scatters off of the AFM tip to the interferometer. **(d)** Mathematical representation of the elliptical fringes of the *roundtrip* contribution in presence of BP. *a* and *b* represent the distance from *m = 0* fringe to *m$^{th}$* fringe along the major and minor axes of the *m$^{th}$* ellipse, respectively. Anisotropy ($\alpha$) is given by $\alpha = \frac{b}{a}$. Ellipticity of the fringe increases with *m*. **(e-h)** Real parts of *roundtrip* contribution of s-SNOM image in Figure S3b for 40 nm h-BN/40 nm BP heterostructure at varying frequencies. Dashed-red circle in **(e)** represents *0$^{th}$ (m = 0)* fringe and the dashed-red ellipse represents *2$^{nd}$ (m = 2)* fringe. *a* and *b* are the distances from the *0$^{th}$* fringe to the major and minor axes of the *2$^{nd}$* fringe, respectively. Arrow represents the direction of the electric field of the illuminating beam. **(i-l)** Real parts of *roundtrip* contribution of s-SNOM image in Figure S3d for 40 nm h-BN/250 nm BP heterostructure at varying frequencies. Number of fringes in the h-BN/BP disc increases with frequency implying increased confinement of the phonon polaritons in h-BN/BP heterostructure. At a given frequency, the mode confinement is larger for thicker BP. **(m)** Effective indices in both 40 nm h-BN/40 nm BP and 40 nm h-BN/250 nm BP heterostructures increases with increasing frequencies implying higher mode confinement as evident in **(e-l)**. Mode confinement is higher for thicker BP which demonstrates the substrate's thickness dependent confinement. On the other hand, the effective indices along armchair axis are larger than the zigzag axis for both the heterostructures. The effective index contrast is lower for 40 nm h-BN/250 nm BP heterostructure implying a lower anisotropy in comparison to 40 nm h-BN/40 nm BP heterostructure. There is an excellent agreement between theory and experiments without the need of fitting any of the parameters of the structure. **(n)** Anisotropy ($\alpha = \frac{b}{a}$) of the phonon polaritons at varying frequencies. In the frequency range 1405-1440 cm$^{-1}$, the anisotropy monotonically decreases for 40 nm h-BN/250 nm BP heterostructure. The anisotropy is lower in comparison to the ratio of refractive indices of BP in the x-y plane shown by dashed-purple line.

Notably, the anisotropy of 40 nm h-BN/40 nm BP heterostructure peaks at 1420 cm$^{-1}$ with a maximum value of $\alpha_{max} = 1.25$. The anisotropy values are much higher in comparison to the ratio of refractive indices of BP in the x-y plane, i.e., $\alpha_{max} > \alpha'$. Thus, the phonon polaritons in h-BN act as a means of enhancing the in-plane optical anisotropy of h-BN/BP heterostructures. Scale bars are 2 $\mu m$.

| vdW Material | Polariton Type | Spectra Range |
|---|---|---|
| h-BN | Phonon Polaritons | 780-830 cm$^{-1}$, 1370-1610 cm$^{-1}$ |
| Undoped BP | (No Polaritons) | - |
| Doped/Gated BP | Anisotropic Plasmon Polaritons | ~ 200-4000 cm$^{-1}$ |
| h-BN/BP Heterostructure | Anisotropic Phonon Polaritons | 780-830 cm$^{-1}$, 1370-1610 cm$^{-1}$ |

**Table 1.** Overview and spectral ranges in which different types of polaritons exist in h-BN, BP and their heterostructure.

Supplementary Materials for **"Engineering Phonon Polaritons in van der Waals Heterostructures to Enhance In-Plane Optical Anisotropy"**

**Authors:** Kundan Chaudhary[1‡], Michele Tamagnone[1‡], Mehdi Rezaee[1,2‡], D. Kwabena Bediako[3], Antonio Ambrosio[4], Philip Kim[3], Federico Capasso[1*]

**Affiliations:**

[1]Harvard John A. Paulson School of Engineering and Applied Sciences, Harvard University, Cambridge, MA 02138, USA.
[2]Department of Electrical Engineering, Howard University, Washington, DC 20059, USA.
[3]Department of Physics, Harvard University, Cambridge, MA 02138, USA.
[4]Center for Nanoscale Systems, Harvard University, Cambridge, MA 02138, USA.

*Correspondence to: capasso@seas.harvard.edu

‡These authors contributed equally to this work.

**Supplementary Method 1: Separation of s-SNOM contributions**

In a sample with an arbitrary shape, the *complex* contributions cannot be separated easily, but given the circular symmetry of the sample, they can be separated in the following way. First, the *material* contribution is constant over the disc, and changes when scanning the substrate outside the disc. Hence it can be retrieved by taking an average of the complex image in an area inside the disc. In fact, the fringes are averaged out, and the mean complex value is the value of the *material* contribution of the disc. The obtained complex value is subtracted from the complex image, and this allows to better distinguish the polaritons in the disc. After the subtraction, the substrate area around the disc takes values much higher than the polariton contributions, so it is set to zero in the images in the paper by masking all the points outside the circle.

The remaining complex-valued fields are the summation of the *direct* and the *roundtrip* contributions. The *direct* contribution is a map of the actual fields that are launched by the edges upon illumination (except for a phase distortion due to scanning explained in ref. 6). Because the mode is transverse magnetic, only the portions of the edge which are orthogonal to the incident polarization will launch the mode efficiently. For example, if the incident light is p-polarized, only the top and bottom part (with respect to the illumination direction) of the edge will launch the mode. In addition, these two segments will launch the modes with opposite phases. Therefore, the *direct* contribution has an odd symmetry with respect to a 180° rotation of the image. On the contrary, the *roundtrip* contribution has an even symmetry with respect to the same transformation. This is because the sample itself is invariant with respect to a 180° rotation and the roundtrip contribution does not depend on the incident polarization, but only on the position of the tip relative to the sample. Hence, to separate the two contributions, the image is first rotated by 180° and then the original and the rotated images are summed together to obtain the *roundtrip* contribution. Similarly, their subtraction gives the *direct* contribution.

The use of the *roundtrip* contribution for our measurement has several advantages. First, it is not affected by the relative orientation of the light polarization and the axis of BP. On the contrary, the *direct* contribution is sensitive, and any misalignment leads to the spiral pattern shown in the main paper, which is of difficult interpretation. Second, the intensity of the *roundtrip* contribution decays faster with increasing distance from the edge, also allowing for an easier interpretation of the fringes. Last, the *direct* contribution phase is affected by the inclination of the incident wave, which requires a correction (*6*). The *roundtrip* contribution suffers from no such effect.

Finally, it is important to notice that the separation approach described above works assuming that the sample has a perfect circular symmetry, so in practice the irregularities in the actual sample might slightly distort the two reconstructed contributions. However, the excellent match with the theoretical predictions indicates that this effect is negligible here.

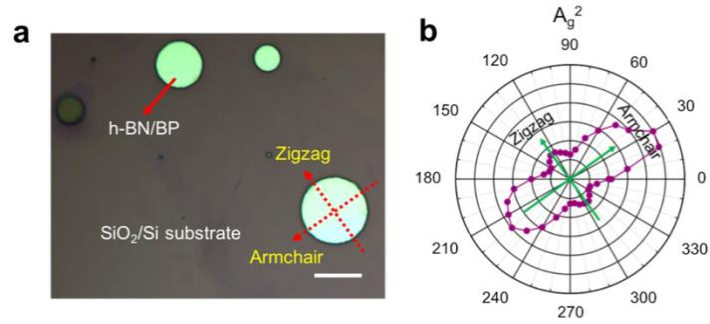

**Figure S1. (a)** Optical image of h-BN/BP heterostructures on $SiO_2$/Si substrate. **(b)** Armchair and zigzag crystal axes are determined using angle-resolved polarized Raman (ARPR) spectroscopy. Intensities of $A_g^2$ along $2\pi$ rotation is represented in the polar plot. Intensities along 30° and 210° are the highest, representing the fast axis, i.e., armchair orientation. Similarly, $A_g^2$ intensities along 120° and 300° are lowest implying the other crystal axis, i.e., zigzag orientation. Scale bar is 10 $\mu m$.

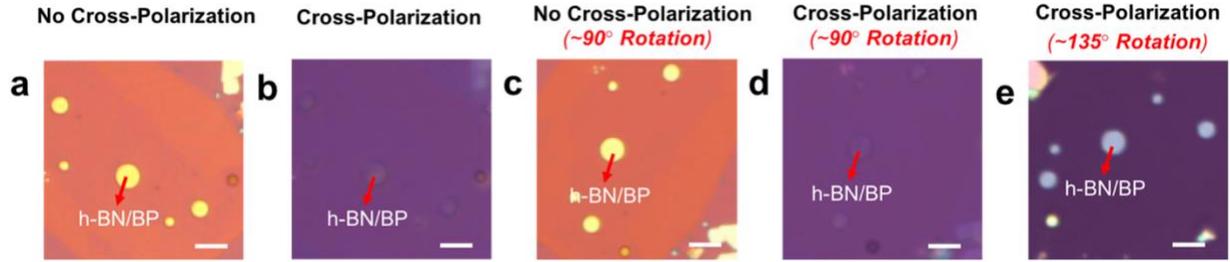

**Figure S2. (a-e)** Optical images (in reflection mode) of h-BN/BP heterostructure discs on SiO$_2$/Si substrate patterned using EBL. Cross-polarization method is used for determining the crystal axes of BP. In **(a-b)** the sample is oriented in such a way that one of the crystal axes is parallel to the polarization of light through the polarizer. With no cross-polarizer **(a)** in the system, the optical image is brighter and darkens when the cross-polarizer is present **(b)**. Similarly, when the sample is rotated by ~90° **(c-d)**, the other crystal axis is now parallel to the polarization of light through the polarizer and the optical image darkens **(d)**. With the crystal axes at an angle with the polarization of light through the polarizer, the h-BN/BP heterostructures are partially reflective and therefore brighter in presence of a cross-polarizer **(e)**. Scale bars are 20 $\mu m$.

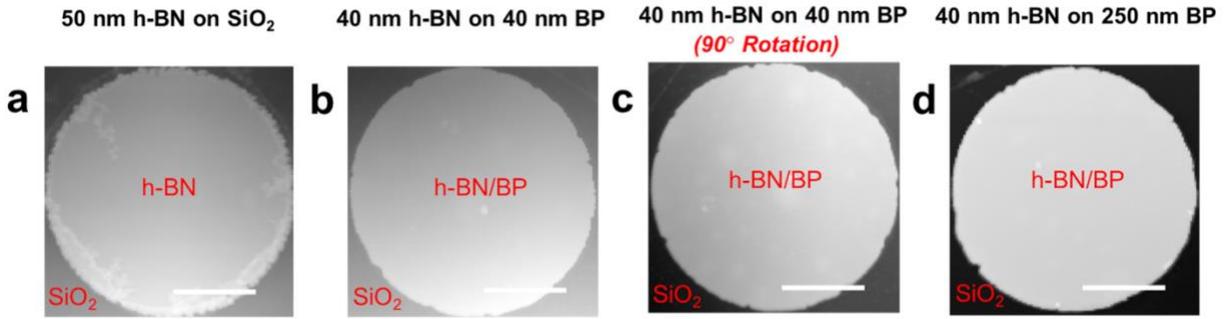

**Figure S3.** AFM images of **(a)** 50 nm h-BN on SiO$_2$/Si **(b)** 40 nm h-BN/40 nm BP heterostructure on SiO$_2$/Si **(c)** 40 nm h-BN/40 nm BP heterostructure on SiO$_2$/Si rotated by 90° **(d)** 40 nm h-BN/250 nm BP heterostructure on SiO$_2$/Si. Scale bars are 2 $\mu m$.

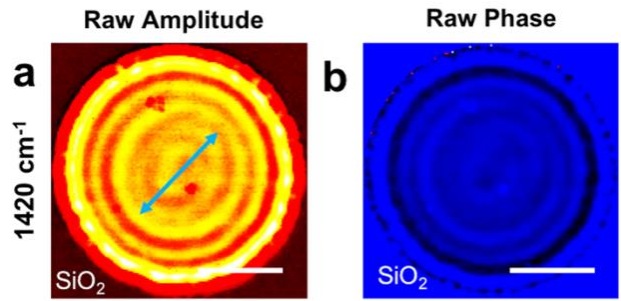

**Figure S4.** Amplitude **(a)** and phase **(b)** of raw near field scan obtained from s-SNOM. These raw images are thereupon used to extract the *direct* and *roundtrip* contributions to further analyze the data. In **(a)**, arrow represents the direction of the electric field of the illuminating beam. Scale bars are 2 μm.

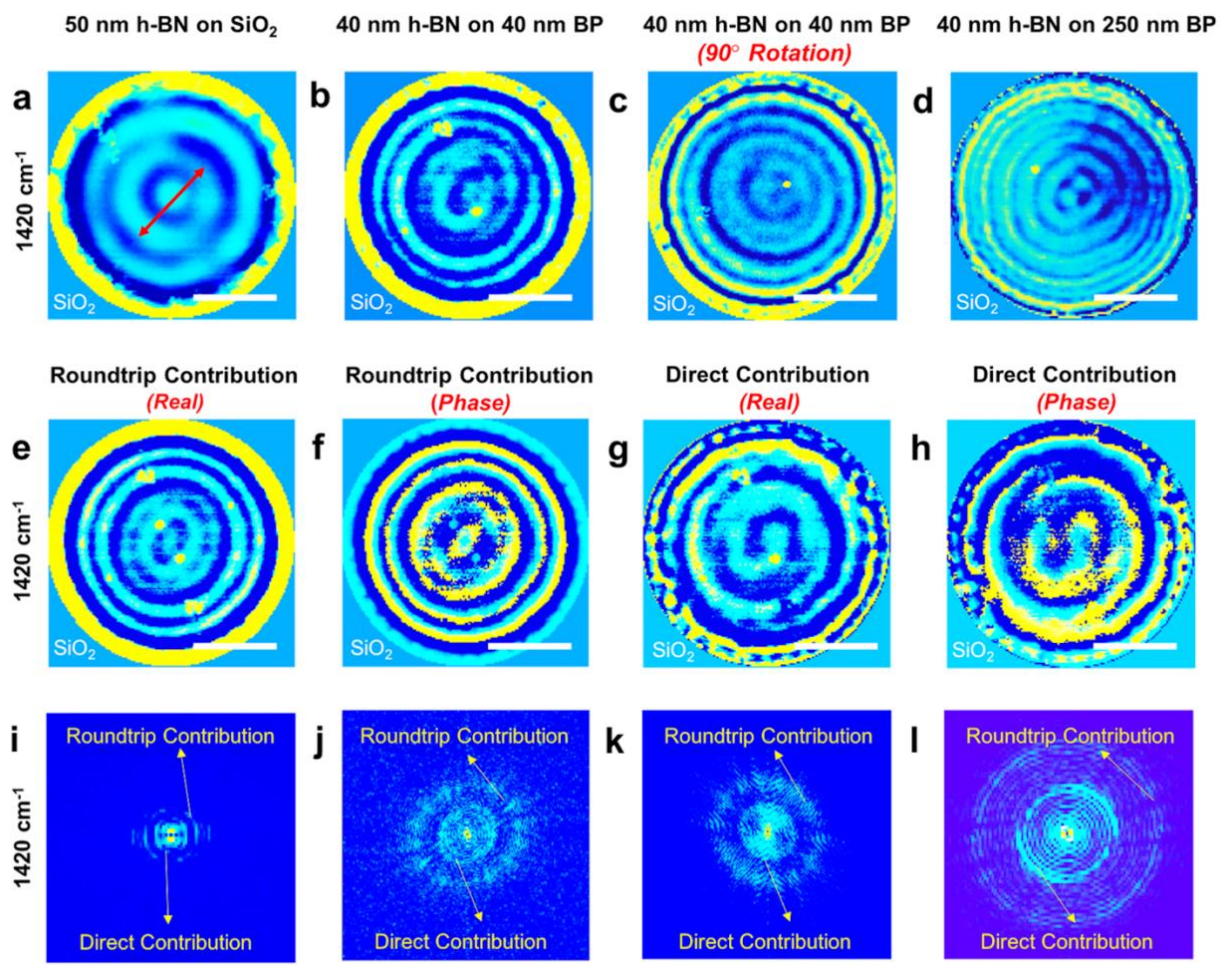

**Figure S5. (a-d)** s-SNOM images at 1420 cm$^{-1}$ after removal of *material* contribution (see Supplementary Method 1) of **(a)** 50 nm h-BN on SiO$_2$/Si **(b)** 40 nm h-BN/40 nm BP on SiO$_2$/Si **(c)** 40 nm h-BN/40 nm BP on SiO$_2$/Si rotated by 90° **(d)** 40 nm h-BN/250 nm BP on SiO$_2$/Si. In **(a)** the fringes are circular as expected for the phonon polariton propagation along an isotropic substrate, i.e., SiO$_2$, but the fringes become approximately elliptical for the cases of **(b-d)** in presence of BP. The anisotropy is invariant with rotation of sample by 90° as evident from **(b-c)** and decreases with increasing thickness of BP in **(d)**. However, 40 nm h-BN/250 nm BP results in high confinement of phonon polaritons given by increased number of fringes within the disc in comparison to the cases in **(a-c)**. Arrow in **(a)** represents the direction of the electric field of the illuminating beam. **(e)** Real part of the *roundtrip* contribution shows elliptical fringes with increased ellipticity of the fringes towards the center of h-BN/BP heterostructure. **(f)** Phase part of the *roundtrip* contribution exhibiting elliptical fringes as shown for real part in **(e)**. **(g-h)** Real and phase parts of *direct* contribution of s-SNOM image in **(b)**, respectively. Such spiral fringes can arise from multiple factors such as rough edges of the h-BN/BP heterostructure. **(i-l)** Fast Fourier Transform (FFT) images of s-SNOM images in **(a-d)** showing a circle for h-BN on $SiO_2$, ellipse for **(b)**, 90° rotated ellipse for **(c)**, and negligible ellipticity for **(d)**. The FFT images show both the *roundtrip* and *direct* contributions given by outer and inner circle/ellipse, respectively. Scale bars are 2 $\mu m$.

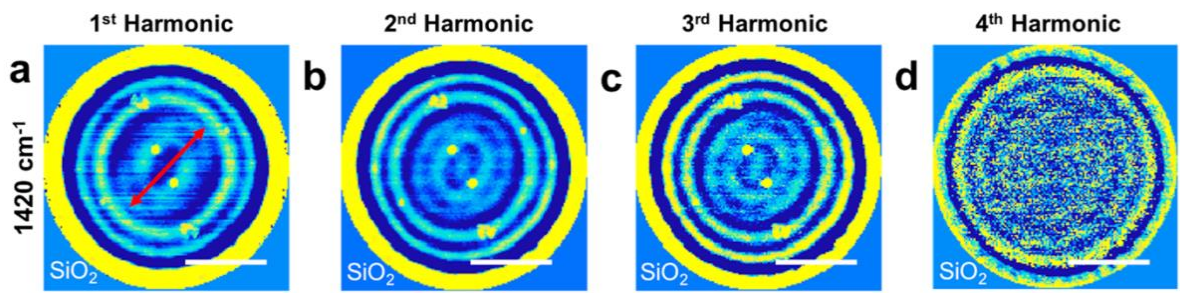

**Figure S6.** s-SNOM images of the real part of the *roundtrip* contribution at different pseudo-heterodyne harmonics. Lower harmonics **(a-c)**, i.e., $n \leq 3$, have stronger near-field signals whereas, the signals are weaker for higher harmonics **(d)**, i.e., $n \geq 4$. The fringe contrast, however, is maximized for the third harmonic. In **(a)**, arrow represents the direction of the electric field of the illuminating beam. Scale bars are $2~\mu m$.

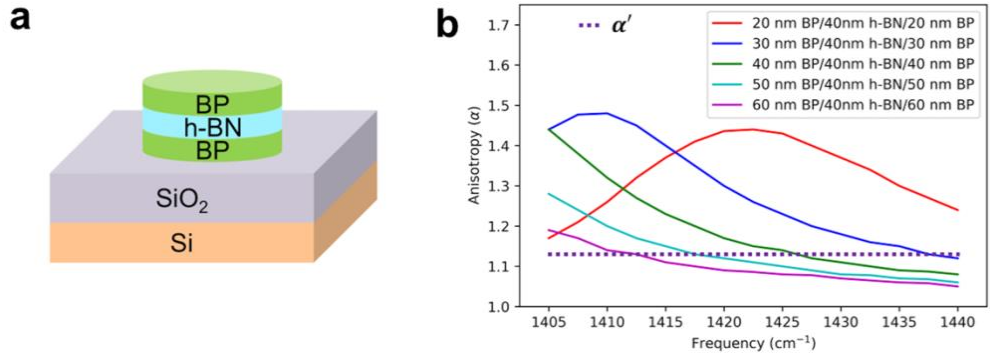

**Figure S7. (a)** Schematic illustration of BP/h-BN/BP heterostructure. **(b)** Calculated in-plane anisotropy values of the phonon polaritons for varying thickness of BP in BP/h-BN/BP heterostructure. The dashed-purple line represents the ratio of refractive indices of BP in the x-y plane. The calculated in-plane anisotropy of BP/h-BN/BP heterostructure can reach up to $\alpha_{max} = 1.5$.